\documentclass[conference]{IEEEtran}
\IEEEoverridecommandlockouts
\usepackage{cite}
\usepackage{amsmath,amssymb,amsfonts}
\usepackage{algorithmic}
\usepackage{graphicx}
\usepackage{textcomp}
\usepackage{xcolor}
\usepackage{float}
\usepackage{url}
\def\BibTeX{{\rm B\kern-.05em{\sc i\kern-.025em b}\kern-.08em
    T\kern-.1667em\lower.7ex\hbox{E}\kern-.125emX}}
\begin{document}

\title{An Open-Source Framework for Rapid Validation of Scientific ASICs
\thanks{This manuscript has been authored by Fermi Research Alliance, LLC under Contract No. DE-AC02-07CH11359 with the U.S. Department of Energy, Office of Science, Office of High Energy Physics. 

This work was funded by the DOE Office of Science Research Program for Microelectronics Codesign through the project “Hybrid Cryogenic Detector Architectures for Sensing and Edge Computing enabled by new Fabrication Processes” (LAB 21-2491).

A. Quinn is with the Fermi National Accelerator Laboratory, Pine \& Kirk St, Batavia, IL 60510
(e-mail: \protect\url{aquinn@fnal.gov}).}
}

\author{\IEEEauthorblockN{Adam Quinn}
\IEEEauthorblockA{\textit{Microelectronics Division} \\
\textit{Fermi National Accelerator Laboratory}\\
Batavia, IL \\
aquinn@fnal.gov}

}

\maketitle

\begin{abstract}
Spacely is an open-source framework for the post-silicon validation of analog, digital, and mixed-signal ASICs (Application-Specific Integrated Circuits) which maximizes the reuse of hardware and software, reducing the time taken to achieve meaningful test results. Spacely specifically addresses the needs of small, flexible ASIC design teams commonly found in academia or research institutions which benefit most from sharing the overhead of test stand creation between many unique ASIC designs. Spacely is a set of software, firmware, and design practices. It targets two primary hardware platforms (NI-PXI and Caribou) as well as offering extensible support for bench instruments. Spacely provides a high-level Python interface to all test hardware for accessibility, while also giving more sophisticated teams the opportunity to integrate custom test firmware. The design principles of Spacely are presented, along with a demonstrative example of using Spacely to test a pixel detector readout ASIC.
\end{abstract}


\section{Introduction}

Modern mixed-signal Application-Specific Integrated Circuits (ASICs) may contain dozens of co-designed analog and digital blocks which interoperate through complex, high-speed interfaces which cannot be directly observed from the outside. They may be extensively reconfigurable and contain millions of bits of unique state. Validating the operation of these devices and measuring their performance is inherently challenging, and often requires the cooperation of experts in a range of domains from noise analysis to digital synthesis, as well as the use of expensive high-performance test instrumentation \cite{rf_testing_challenges}. This challenge is felt most acutely in academic and research environments where the ASICs under test probe the edge of technical possibility, and may contain specialized functionality that does not fit traditional test regimes, such as operation in extreme environments or novel sensing methods. Research teams also lack the scale and resources of large IC design companies, and often work on a variety of disparate chip designs instead of making incremental improvements on a few established product lines, making it cost-prohibitive to develop extensive customized test solutions.

In such a context, obtaining ASIC test results in a reasonable timeframe requires streamlining and generalizing the test process to the greatest extent possible, freeing experts to focus on the unique challenges of their novel design. An open-source validation framework which maximizes reuse of code and hardware has the potential to address this challenge, while being within the financial reach of academic and research institutions. \cite{inequality}

Open-source and reusable architectures for designing chips have become prolific in recent years \cite{chipkit} \cite{Antmicro_2021}, and industrial-quality ASICs have been demonstrated using fully open-source EDA flows like OpenLane \cite{real_open_source_silicon}. Yet when it comes to hardware test automation, ad-hoc in-house solutions and expensive commercial systems are still dominant \cite{BREAD}. Recently, progress has been made in developing open-source data acquisition hardware \cite{BREAD} \cite{SLAC_ATCA} \cite{Caribou_1} \cite{Caribou_2} \cite{opendaq} and software infrastructure \cite{slac_rogue}, however a greater degree of abstraction and rapid reconfigurability is needed to meet the needs of low-volume, quick-turnaround research environments. 

\begin{figure}[htbp]
\centerline{\includegraphics[width=\linewidth]{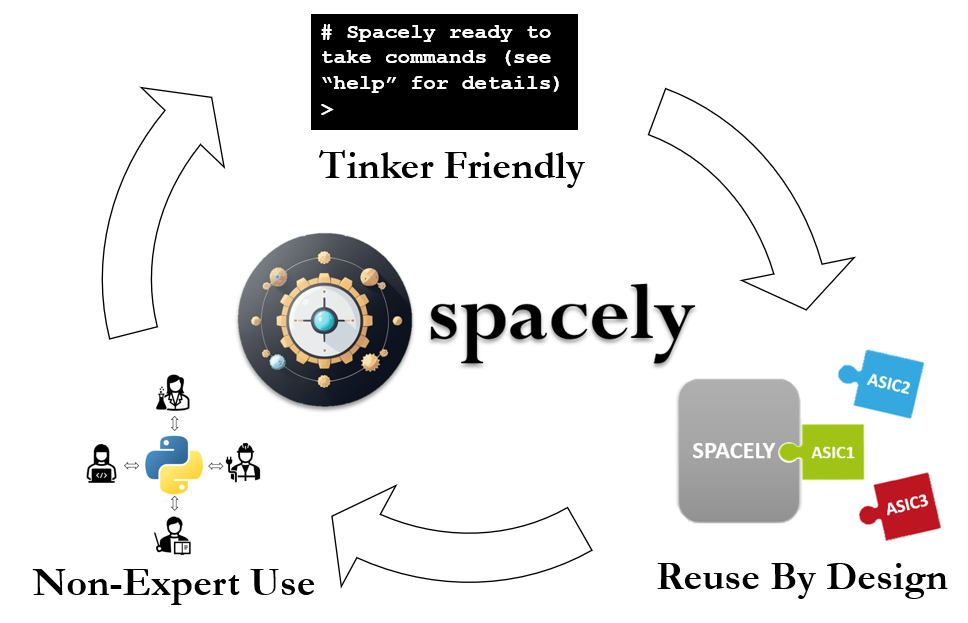}}
\caption{Spacely Design Principles}
\label{fig_spacely_principles}
\end{figure}

\begin{figure*}[htbp]
\centerline{\includegraphics[width=0.8\textwidth]{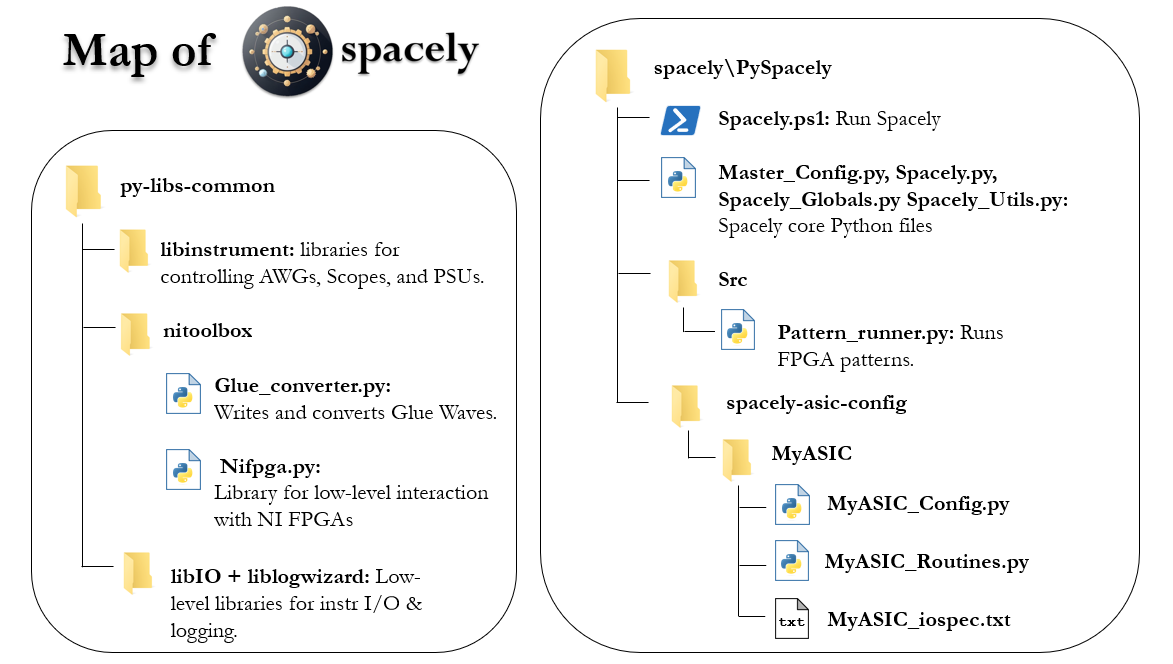}}
\caption{The basic structure of the three Spacely code repositories, and key files found within each. \textbf{spacely-asic-config}, which contains ASIC-specific code, is a submodule of the core \textbf{spacely} repository. Both \textbf{spacely} and \textbf{spacely-asic-config} call library functions from \textbf{py-libs-common}. Bolded filenames in the following sections can be found in this figure.}
\label{fig_spacely_map}
\end{figure*}

Spacely is an open-source framework for validating analog, digital, and mixed-signal scientific ASICs. It can interface to either NI-PXI, a leading commercial data acquisition system, or to Caribou, an open-source data acquisition hardware system also developed in the scientific community. Spacely provides a unified Python environment for setting up instruments, scripting tests, and analyzing data. Development of this framework adheres to a set of design principles (Figure \ref{fig_spacely_principles}):

\begin{itemize}
    \item \textbf{Non-Expert Use:} Any person who can write Python should be able to write ASIC tests with Spacely, and understand tests that have been written by others. This is crucial to prevent information from being lost between engineers who use different tools (i.e. design vs test engineers), and also allows people such as science domain experts and students to be involved in test. 
    \item \textbf{Reuse by Design:} Design-specific code should be limited to the bare minimum and kept in a single location so that the user can "swap in" to a different ASIC while keeping the vast majority of the spacely code base exactly the same.
    \item \textbf{Tinker Friendly:} If the user knows how to perform a task with a snippet of Python, they should be able to write that snippet and run it easily in the Spacely environment. If the task is common enough, Spacely should provide a short shell command to perform it, but the interface should never get in the user's way. (For this reason, Spacely eschews GUIs in favor of shells.)
\end{itemize}

The remaining sections of this paper discuss the core Spacely codebase and workflow for using Spacely (Sections II and III), Spacely integration with NI-PXI (Section IV) and Caribou (Section V), and an example of using Spacely to rapidly validate a scientific ASIC (Section VI).

\section{Spacely Workflow}

The Spacely codebase is divided into three mutually-dependent repositories in order to promote code reuse (see Figure \ref{fig_spacely_map}).

The main \textbf{spacely} repository contains core code for Spacely's interactive shell and for integration with the NI-PXI and Caribou hardware systems described later, along with various utilities for analyzing and manipulating data that is collected from the ASIC under test.

The \textbf{spacely-asic-config} repository, a submodule of \textbf{spacely}, contains a subfolder for every ASIC which is tested using Spacely. This subdirectory contains all code which is specific to that ASIC, such as settings for the values of voltage and current biases, or custom routines written to test a specific block implemented on that ASIC. 

Finally, the \textbf{py-libs-common} repository contains instrument drivers and basic libraries for communication protocols and logging which may have use even beyond Spacely.

Spacely is designed to be accessible to users of varying levels of sophistication, from ASIC/FPGA designers to scientists or students who can write Python but have limited hardware experience. It is possible to fully test many mixed-signal ASICs simply by writing Python routines in the \textbf{spacely-asic-config} area. However, as an open-source project, teams with development experience can also extend \textbf{py-libs-common} to support additional test instruments or develop custom firmware, depending on the primary hardware platform. 

The sections below give an overview of the steps that must be taken to validate most ASICs:

\subsection{Select a Primary Hardware Platform}

Spacely targets two primary hardware platforms which are used to communicate with an ASIC. The details of using each of these platforms is presented in later sections. 

NI-PXI is a proprietary test automation platform developed by National Instruments. In a typical configuration, it consists of a chassis which can host multiple pluggable "modules" including FPGAs, power supplies, and bias generators.

Caribou \cite{Caribou_1} \cite{Caribou_2} is an open-source data acquisition platform developed by a collaboration of universities and research institutions. It is based around a Xilinx SoC with a Control-and-Readout (CaR) board that supplies current and voltage supplies, ADCs, DACs, and pulsers.

While NI-PXI is an established platform with commercial heritage, and may scale more easily to accomodate ASICs with a very high number of I/Os, the Caribou platform offers a much lower upfront cost and zero ongoing license fees, and is generally more customizable, and allowing users with RTL experience to develop custom firmware blocks with industry-standard tools (Vivado). The Caribou platform also has unique features such as high-speed GTx links.

The Spacely conceptual framework and design principles can easily be extended to alternative FPGAs or hardware platforms. However, extending the codebase to be compatible with these platforms is non-trivial. The author has carried out significant work to create bitfiles (for NI-PXI) and OS images (for Caribou) which are made freely available to the Spacely user, but would have to be replicated for another platform.

ASIC carrier PCBs must be designed to fit the I/O specification of the selected hardware system. However, considering the flexibility of both systems, a large degree of PCB design reuse can be expected. 

\subsection{Define an ASIC Configuration}

The first file that should be written when testing an ASIC with Spacely is typically \textbf{spacely-asic-config/MyASIC/MyASIC\_Config.py}. This file describes the test hardware setup in a Python format, and specifies everything that should be initialized when Spacely starts. The format of this file is described in Spacely Documentation. In general, it defines:

\begin{itemize}
    \item A list of test instruments that Spacely will control, along with their appropriate communication interfaces / addresses.
    \item FPGA bitfiles to be used (for the NI-PXI system)
    \item Voltage rails and current biases -- their nominal values, limit values, and order of initialization.
\end{itemize}

\subsection{Write ASIC Test Routines}

The second key configuration file, \textbf{spacely-asic-config/MyASIC/MyASIC\_Routines.py} defines test routines which the user wants to run as Python functions. Any function prefixed by the keyword "ROUTINE\_"  can be called interactively from the main Spacely shell. These routines can consist of arbitrary Python code, making them extremely flexible, however Spacely provides convenient idioms to access instrument methods (for example, changing voltages and currents or taking data from an oscilloscope), send and receive digital waveforms from the ASIC, and process data.

\begin{figure}[htbp]
\centerline{\includegraphics[width=\linewidth]{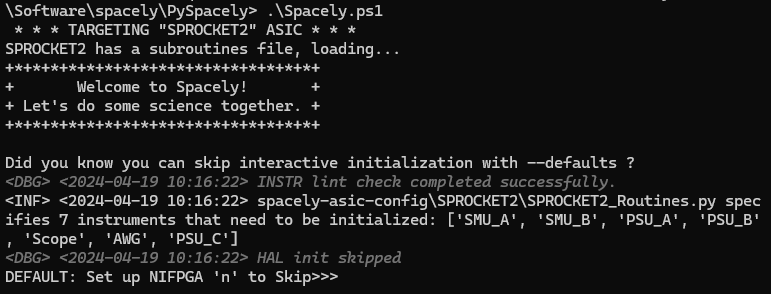}}
\caption{The core Spacely shell. On startup, Spacely loads the configuration and routines written for the target ASIC (in this case "SPROCKET2") and prompts the user to initialize appropriate test instruments. Initialization can also be made automatic with the --defaults flag.}
\label{fig_startup}
\end{figure}

\subsection{Run Tests and Collect Data}
 
When Spacely is run, it references the \textbf{Master\_Config.py} file to determine which ASIC is being targeted. Based on this target specification, it will load Config and Routines files and initialize the test stand accordingly (Figure \ref{fig_startup}). (For example, to load the \textbf{MyASIC\_Config.py} and \textbf{MyASIC\_Routines.py} files, \textbf{Master\_Config.py} would specify \textbf{TARGET="MyASIC"}). This configuration switch allows a test engineer to rapidly switch between targeting different ASICs as necessary. 

After initialization, the main Spacely shell is opened. This shell provides idioms for quickly running routines defined in the loaded Routines file or monitoring system voltages. It also functions as a Python interpreter, allowing the user to, for example, change voltage supplies on-the-fly for quick testing. 

When the Spacely shell closes, all test instruments and supplies that were initialized are automatically turned off. 

\begin{figure*}[htbp]
\centerline{\includegraphics[width=0.9\textwidth]{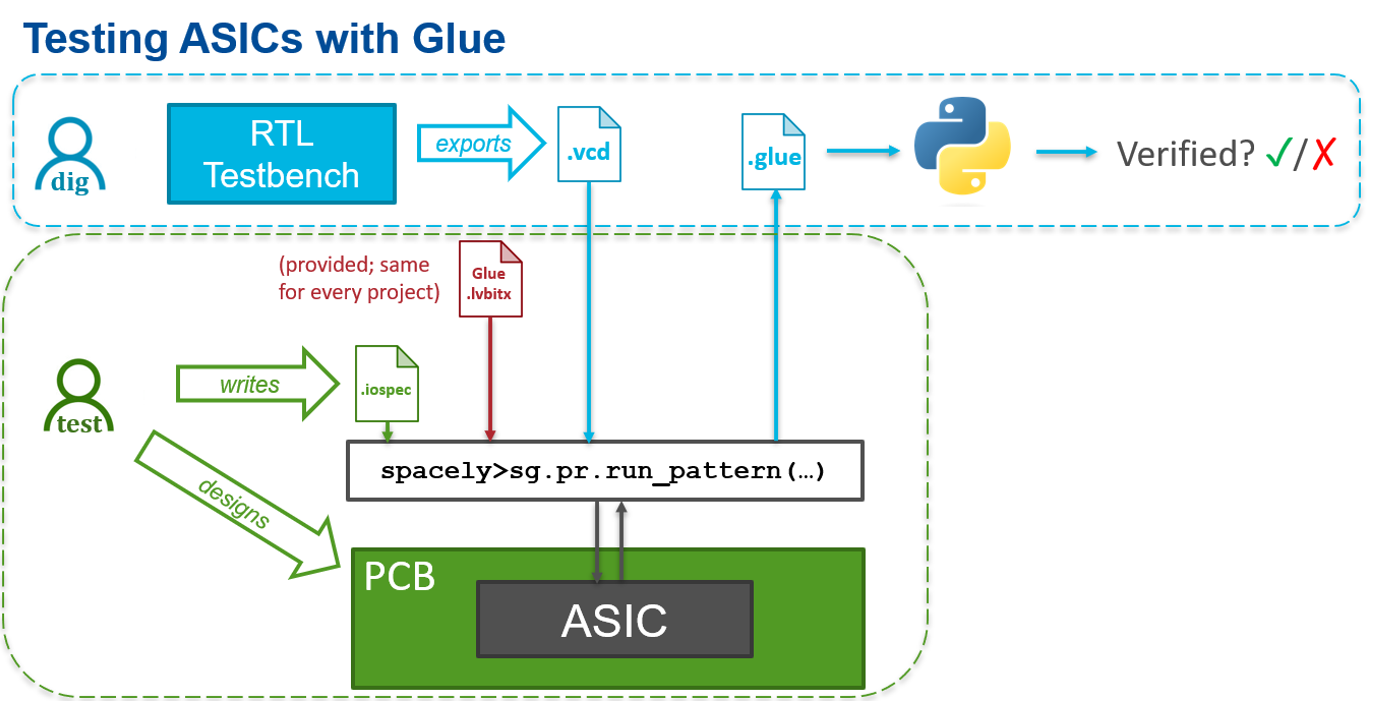}}
\caption{Verilog-to-Lab flow for using Spacely to test a digital / mixed-signal ASIC. Responsibility is efficiently divided between design (blue) and test (green) engineers. sg.pr.run\_pattern() is a Spacely provided function which runs a Glue wave on an ASIC.}
\label{fig_glue}
\end{figure*}

\section{Key Features}

\subsection{Verilog-to-Lab Flow}

Spacely extends the principle of Reuse By Design beyond the realm of test engineering to ASIC verification. A gap exists between the tools commonly used for digital ASIC verification or emulation (SystemVerilog testbenches, UVM, or Cocotb) and those used for test (NI-PXI, logic analyzers, bespoke FPGA systems), which can create enormous inefficiency and the potential for errors if the details of an ASIC's operation and its test conditions are mistranslated between design and test groups.

Spacely partially bridges these two domains by using \textit{Glue Waves}. A Glue Wave is a simple, compressed format for representing arbitrary digital patterns as a series of integers, in which the binary bits of each integer represent the values of particular I/Os at a given timestamp. (The name "Glue" is an allusion to glue code, which, in software engineering, allows two subsystems that do not share an API to communicate.) Spacely provides a utility for generating Glue Waves from Value Change Dump (VCD) files, which are commonly generated by logic tools, and it provides arbitrary wave generation firmware for Caribou and NI-PXI that use Glue Waves as both input and output. For example, in a simple serial interface with a clock (SCK), data-in (SDI), and data-out (SDO), a digital designer may export the SCK and SDI waves from their Cocotb testbench, convert the format to a Glue Wave, and send that Glue Wave to the NI-PXI pattern runner. The pattern runner returns a new Glue Wave which contains the original SCK and SDI as well as an SDO wave from sampling the actual ASIC output. The designer can then use tools within Spacely to compare or graphically present the Glue waves, or extract a logical bitstream from the wave to verify the ASIC's correctness. 

This process, illustrated in Figure \ref{fig_glue}, closely involves the digital ASIC designer in test, ensuring specifications are met. The ASIC test engineer may focus on the physical test stand design, which is relatively agnostic to the function of the ASIC. The physical connectivity between the Spacely-driven wave generator and the ASIC is encoded in an additional configuration file \textbf{spacely-asic-config/MyASIC/MyASIC\_iospec.txt} which contains a simple mapping of I/O pin numbers to signal names.

For situations where an RTL testbench is not available for a specific functionality, or when test is driven by a scientist or user rather than a digital engineer, Spacely also provides utilities to generate glue waves (a) Natively within Python, as a dictionary containing time-series values for each signal, or (b) By hand, writing an ASCII file in which signal levels are represented by 1's and 0's. The high productivity and Tinker Friendliness of Python allows for complex changes to digital control patterns to be made on the fly.

One caveat to the Glue Wave framework is that feedback is not possible within a single waveform because the entire waveform is written as a block and cached by the FPGA prior to receiving the ASIC's response. However, in practice this is often not a concern if communication between the ASIC and the test stand can be separated into transactions. For example, a Glue Wave can be generated for a SPI read transaction and sent to the ASIC. Then, based on the values of the data read, Spacely can dynamically generate a new Glue Wave for a SPI write transaction. In the case where tight feedback coupling is necessary, custom firmware must be designed. When using Caribou, Spacely provides a procedure for smoothly integrating this firmware into the test framework, discussed below.

\subsection{Metadata Tagging}

Testing complex mixed-signal ASICs can potentially involve setting dozens or hundreds of variables, from the levels of voltage and current biases to the timing of digital control patterns. Recording the values of these variables and matching them to collected data points is crucial to ensure that results from testing are provable and repeatable. Spacely simplifies this process by providing an object-oriented system for metadata tagging through Experiments and DataFiles. 

An Experiment object represents a test routine executed at a particular moment in time, and metadata attributes can be associated to it with \textbf{Experiment.set("attr",value)}. Each experiment contains a collection of DataFiles, which represent CSV files with arbitrary structure and can also have metadata attributes set by with \textbf{DataFile.set("attr",value)}. In the case where both the experiment and the DataFile have a setting for "attr", the DataFile overrides. 

The recommended practice is to write a high-level test routine which defines the values that each variable will take in a DataFile object, then pass that DataFile object to a low-level subroutine which will retrieve those variable values from the DataFile, carry out the appropriate test, and write results back to the DataFile. The DataFile then serves as an abstract interface between the two function, and so long as this interface is respected (i.e. no extra variables are passed along with the DataFile), the DataFile will contain all the metadata necessary to reproduce the experiment. 

When results are written to file, a folder is created with the name of the Experiment and populated with DataFiles, along with a metadata file which records the values of metadata for each DataFile and for the experiment as a whole.

\subsection{Spacely Micro-Shells}

In order to support quick experiments, Spacely implements a variety of useful commands in the shell. In particular, it implements three "micro-shells" which can be invoked by typing their name in the main shell. Within each of these shells, additional commands are available for specific tasks.

The Analysis Shell (\textbf{ashell}) implements basic plotting (histogram, scatter plot) and data analysis capabilities for DataFiles. 

The Glue Converter Shell (\textbf{gcshell}) provides tools for manipulating Glue waves. The gcshell can be used to plot Glue waves, edit them, change particular bits, or compare waves against each other.

The I/O Shell (\textbf{ioshell}) provides immediate access to all digital I/Os when using the NI-PXI hardware. The user can select individual signals and toggle them manually.

\section{Spacely Integration with NI-PXI}

Spacely integrates with the PXI data acquisition system provided by National Instruments. NI-PXI power supply and source measurement unit modules (PSUs and SMUs) are treated as normal voltage/current supplies within the Spacely framework, and NI FPGAs are used to send arbitrary digital waves to the ASIC under test.

When the NI-PXI system is used by itself, LabView is used to graphically create FPGA firmware, which is then compiled to a bitfile either locally or on NI servers and then flashed to the FPGA. This process requires several LabView licenses and a lot of processing time, and is generally incompatible with all three Spacely design principles.

When using Spacely, a single pre-generated bitfile is provided for each supported FPGA. This bitfile implements a digital arbitrary wave generator, which can be triggered or stopped using a software API. To communicate with the ASIC, the Spacely user first generates a Glue Wave in Python, then sends it to the waveform generator as previously discussed.

\begin{figure}[htbp]
\centerline{\includegraphics[width=\linewidth]{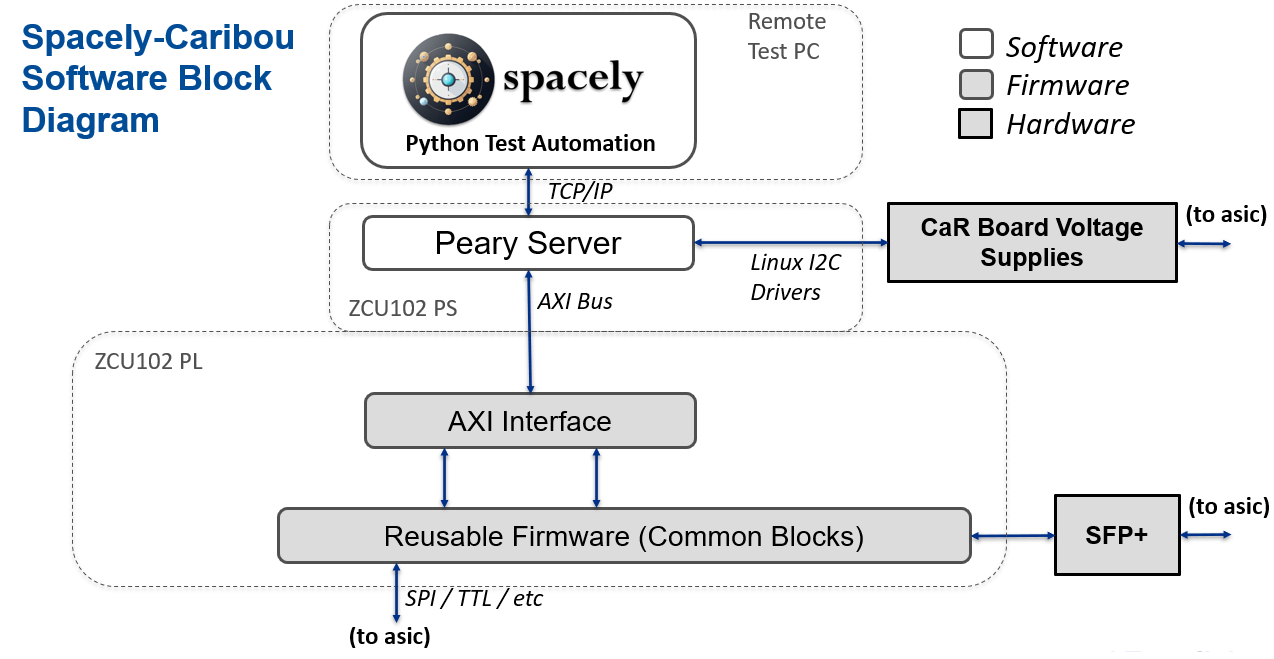}}
\caption{Block diagram of the software and firmware used in a Spacely-Caribou integration.}
\label{fig_caribou}
\end{figure}

\section{Spacely Integration with Caribou}

Spacely integrates with the Caribou open-source data acquisition platform. The primary components of Caribou include a ZCU102 Xilinx SoC comprising an programmable logic fabric (PL) and a processor (PS) running Linux, as well as the Control-and-Readout (CaR) board, a PCB that provides an array of ADCs, DACs, and other peripheral devices controlled from the PS over I2C.

Caribou can be used independently of Spacely. In this scenario, a user flashes an image to the PL, then compiles an instance of Peary, the Caribou system's control software (which is also open-source), on the PS. Peary communicates with the firmware over an AXI bus and controls the CaR board devices using standard Linux I2C drivers. When integrating with Spacely, the process is largely the same, with Spacely sending commands to Peary over a TCP/IP interface as shown in Figure \ref{fig_caribou}, providing a higher level of abstraction and allowing integration with other test instruments. 

Spacely and the ZCU102 can also be used together, independently of the Control-and-Readout board. In this scenario, the Caribou ADC and DAC channels are not available, but the digital I/Os of the ZCU102 may be accessed with a breakout mezzanine.

Unlike NI PXI, there is not one standard firmware bitstream provided for Spacely-Caribou. Instead, Spacely-Caribou specifies a firmware architecture, in which all firmware blocks are controlled from a single memory-mapped AXI bus. A repository of common firmware blocks, \textbf{spacely-caribou-common-blocks} which all have an AXI peripheral interface, is under development and provided to the community. Xilinx also provides a rich library of AXI-controlled IP, including AXI GPIOs and AXI timers. By leveraging these repositories and Vivado's automatic connection assistance, a Spacely-Caribou firmware image can be built simply by dragging and dropping desired blocks into a block diagram, connecting pins, and synthesizing. Of course, expert users can also contribute their own firmware blocks.

\section{Spacely in Practice}

Spacely has been used to test ASICs at Fermilab for approximately a year. One notable example is SPROCKET2, a complex mixed-signal ASIC containing an experimental ADC. During initial investigation, the ADC displayed substantial non-linearity in one gain region. Simulation suggested that this could be due to insufficient settling time in one phase of the ADC's operation. However, the ADC was controlled by a complex digital pattern involving ten distinct signals, so it was unclear where the issue existed. 

For this ASIC, the NI-PXI hardware system was used, and the digital control pattern was generated programmatically as a Python dictionary (eventually converted into a Glue wave). To debug this issue, the pattern generation was first parameterized using variable to represent the length of each particular phase, for example \textit{flip1} as shown in Figure \ref{fig_ex1}:

\begin{figure}[H]
\centerline{\includegraphics[width=\linewidth]{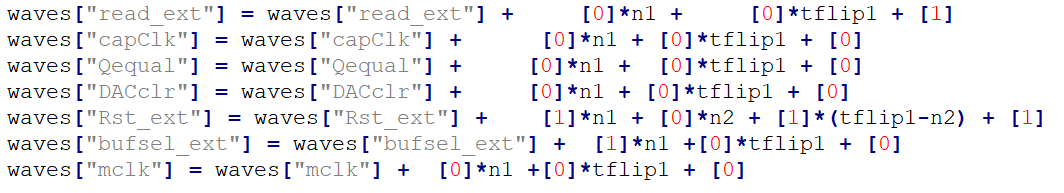}}
\caption{A code snippet from the function that generates the ADC control pattern for SPROCKET2.}
\label{fig_ex1}
\end{figure}

A Spacely Routine was written to vary the lengths of these phases, and record a full transfer function of the ADC for each variation. For each variation, a new DataFile was created, and the value of the swept parameter (for example \textit{tflip1}) was set as a piece of metadata in that DataFile as shown in Figure \ref{fig_ex2}:

\begin{figure}[H]
\centerline{\includegraphics[width=\linewidth]{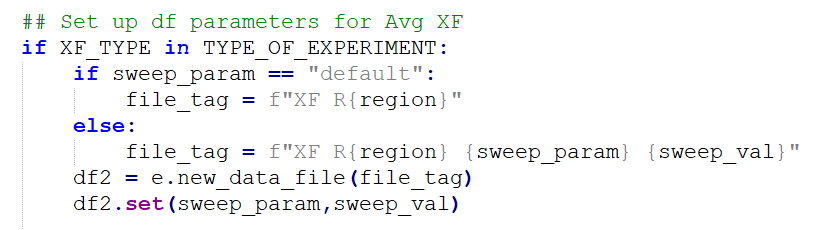}}
\caption{A code snippet from a high-level Spacely routine which creates DataFiles corresponding to each variation of the selected parameters. }
\label{fig_ex2}
\end{figure}

\begin{figure}[H]
\centerline{\includegraphics[width=\linewidth]{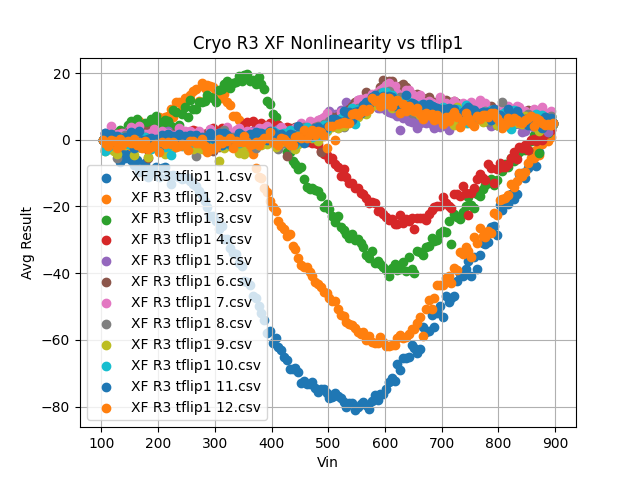}}
\caption{A graph generated using \textbf{ashell} of the transfer function nonlinearity (ADC codes per mV input) versus \textit{tflip1} values from 1 to 12 clock cycles. "Cryo R3 XF" refers to the transfer function of Gain Region 3, measured at cryogenic temperature.}
\label{fig_ex3}
\end{figure}

Ultimately, Spacely collected data from 40 transfer functions with $\approx1,000$ data points each, while the author of this paper went for a coffee that he very much deserved. When the data collection was complete, the data points were imported into \textbf{spacely\>ashell}. The ashell command \textbf{cancel\_linear} was used to remove the linear term from the transfer function, making the nonlinear terms more obvious, and the \textbf{mscatter} command was used to create a comparative scatter plot (Figure \ref{fig_ex3}) showing the difference as \textit{tflip1} is swept from 1 to 12. 

The results (collected and analyzed within Spacely) clearly confirmed that the length of the \textit{flip1} phase has a large effect on the observed nonlinearity. With \textit{tflip1} greater than 5 clock cycles, the primary nonlinear term vanishes, though a smaller, unrelated nonlinearity persists.

\section{Conclusion}

Spacely provides an open-source, test automation platform for analog, digital, and mixed-signal ASICs that simplifies the test and analysis process for both expert and non-expert users.

Within the first year of its use at Fermilab, it has been used in the test of eight different ASIC designs with diverging interfaces and test specifications. The use of Spacely has alleviated a testing resource bottleneck by accelerating the time to get meaningful results: for several ASICs, key digital functionality was demonstrated within the first day of testing after receiving the chip. Work is currently underway to expand the number of Spacely-Caribou test stands in use, and to extend Spacely libraries to control a larger range of test instruments. 

Spacely is a growing open-source project that benefits from community engagement. Organizations interested in using Spacely in their labs, or in further developing Spacely code, can find the source code on Github at https://github.com/SpacelyProject/spacely and documentation at https://github.com/SpacelyProject/spacely-docs. Spacely is released under an Apache 2.0 License. Inquiries can be directed to the developers at: SpacelyDevelopers@fnal.gov.

\bibliographystyle{IEEEtran}
\bibliography{main.bib}

\end{document}